\documentclass{PoS}

\usepackage{cite}

\def\der{{\rm d}}

\def\Q2{\left(Q^{2}\right)}

\def\d{{\rm d}}

\def\l({\left(}
\def\r){\right)}

\title{NNLO Antenna Subtraction with One Hadronic Initial State}

\ShortTitle{NNLO Antenna Subtraction with One Hadronic Initial State}

\author{Alejandro Daleo, Aude Gehrmann-De Ridder\\
        Institute for Theoretical Physics, ETH Z\"urich\\
        E-mail: \email{adaleo@phys.ethz.ch}, \email{gehra@phys.ethz.ch}}

\author{Thomas Gehrmann, \speaker{Gionata Luisoni}\\
        Institut f\"ur Theoretische Physik, Universit\"at Z\"urich\\
        E-mail: \email{thomas.gehrmann@physik.uzh.ch}, \email{luisonig@physik.uzh.ch}}

\abstract{In this talk we present the extension of the antenna subtraction method to include
initial states containing one hadron at NNLO. We sketch the requirements for the different necessary subtraction terms, and we explain how the 
antenna functions are integrated over the appropriate phase space by reducing the integrals to a small set of master integrals.
Where applicable, our results for the integrated antennae were cross-checked against the known NNLO coefficient functions for deep inelastic scattering processes.}

\FullConference{RADCOR 2009 - 9th International Symposium on Radiative Corrections (Applications of Quantum Field Theory to Phenomenology) ,\\
		 October 25 - 30 2009\\
		 Ascona, Switzerland}

\begin{document}

\section{Introduction}
Final states containing hadronic jets are produced at large rates at high energy particle colliders. Owing to their large production cross sections, various jet observables can be measured to a high statistical accuracy. However, experimental data on these observables are often so precise that meaningful precision studies must rely on theoretical predictions that in perturbative QCD requires corrections at next-to-next-to-leading order (NNLO). 

At NLO and NNLO the contribution from real and virtual corrections are separately divergent and, while infrared singularities from purely virtual corrections are obtained immediately after integration over the loop momenta, their extraction from purely real or mixed real-virtual contributions is much more involved. In the latter case the singularities become explicit only after integrating the matrix element over the appropriate phase space. Since the integration is in most of the cases not feasible analytically, one need to extract infrared divergencies at the integrand level and construct subtraction terms which should satisfy the following two conditions: (A) they should approximate the real radiation matrix element in all singular limits, and (B) they should be sufficiently simple to be integrated analytically over a section of the phase space that encompasses all regions corresponding to singular configurations.
In the contribution to an $m$-jet cross section at NLO and NNLO the subtraction terms are added and subtracted in the following way:
\begin{eqnarray}
\der\sigma_{\rm NLO}&=&\int_{\der\Phi_{m+1}}\left(\der\sigma_{\rm NLO}^{\rm R}-\der\sigma_{\rm NLO}^{\rm S}\right)+\left[\int_{\der\Phi_{m+1}}\der\sigma_{\rm NLO}^{\rm S}+\int_{\der\Phi_{m}}\der\sigma_{\rm NLO}^{V,1}\right]\label{eq:nlosub}\\
\der\sigma_{\rm NNLO}&=&\int_{\der\Phi_{m+2}}\left(\der\sigma_{\rm NNLO}^{\rm R}-{\der\sigma_{\rm NNLO}^{\rm S}}\right) +\int_{\der\Phi_{m+1}}\left(\der\sigma_{\rm NNLO}^{\rm V,1}-{\der\sigma_{\rm NNLO}^{\rm VS,1}}\right)\nonumber\\ 
&&\int_{\der\Phi_{m+2}}\der\sigma_{\rm NNLO}^{\rm S}+\int_{\der\Phi_{m+1}}\der\sigma_{\rm NNLO}^{\rm VS,1}+
\int_{\der\Phi_{m}}\der\sigma_{\rm NNLO}^{\rm V,2}\label{eq:nnlosub}\,,
\end{eqnarray}
where $\der\sigma_{i}^{\rm R}$ is the real radiation contribution, $\der\sigma_{i}^{\rm S}$ and $\der\sigma_{\rm NNLO}^{\rm VS,1}$ the subtraction terms of the real and real-virtual contributions respectively and $\der\sigma_{i}^{\rm V,n}$ is the virtual contribution at $n$ loop ($i=$NLO, NNLO).

In the last decades several subtraction methods were developed at NLO~\cite{kunszt,cs,ant,nlosub}
and NNLO~\cite{Kosower:2002su,nnlosub2,nnlosub3,nnlosub4,nnlosub5,nnlosub6}. One of these methods is the so-called antenna subtraction, which was derived at NNLO in~\cite{ourant} for partons only in the final state. The antenna subtraction formalism
constructs the subtraction terms from antenna functions. Each antenna function encapsulates all singular limits due to the
emission of one or two unresolved partons between two colour-connected hard radiator partons. This construction exploits the  universal factorization of matrix elements and phase space in all unresolved limits. The antenna functions are derived systematically from physical matrix elements~\cite{our2j}. For processes with initial-state partons, antenna subtraction has been fully worked out only to NLO so far~\cite{hadant}. In this case, one encounters two new types of antenna functions, initial-final antenna function with one radiator parton
in the initial state, and initial-initial antenna functions with both radiator partons in the initial state.
In this talk we present very recent results on the derivation of all NNLO initial-final antenna functions and their integration over the appropriately factorized phase space. Finally we show how our results could be cross-checked for many cases against the known NNLO coefficient functions for deep inelastic scattering.
The initial-final antenna functions form part of the full set of antenna functions needed for NNLO calculations of hadron collider processes, and are, together with the already known final-final antenna subtraction terms, sufficient for NNLO calculations of jet observables in deeply inelastic lepton-hadron scattering.

\section{Initial final antenna subtraction at NNLO}
In the case of one parton in the initial state the subtraction term at NLO has the following form
\begin{eqnarray}\label{eq:subif}
&&\d\hat{\sigma}^{S,(if)}(p,r)={\cal N}\sum_{m+1}\d\Phi_{m+1}(k_1,\dots,k_{m+1};p,r)
  \,\frac{1}{S_{m+1}}\nonumber\\
&&\times\sum_{j} X^{0}_{i,jk}
  \left|{\cal M}_m(k_1,\dots,K_{K},\dots,k_{m+1};x p,r)\right|^2\,
  J^{(m)}_{m}(k_1,\dots,K_{K},\dots,k_{m+1})\,,
\end{eqnarray}
where $i$ labels the hard radiator with momentum $p$ in the initial state. The additional momentum $r$ stands for the momentum of the second
incoming particle which can either be colored or colorless. This contribution has to be appropriately convoluted with the parton distribution function $f_i$. The tree antenna $X^{0}_{i,jk}$, depending only on the original momenta $p$, $k_j$ and $k_k$, contains all the configurations in which parton $j$ becomes unresolved. The $m$-parton amplitude depends only on redefined on-shell momenta $k_1,\dots,K_{K},\dots,$ and on the momentum fraction $x$. The jet function, $J^{(m)}_{m}$, in (\ref{eq:subif}) depends on the momenta $k_j$ and $k_k$ only through $K_K$.

As eq.~(\ref{eq:nnlosub}) shows, at NNLO two types of contributions to $m$-jet observables require subtraction: the tree-level  $m+2$ parton matrix elements (where one or two partons can become unresolved), and the one-loop $m+1$ parton matrix elements (where one parton can become
unresolved). In ${\rm d}\sigma^{S}_{NNLO}$, we have to distinguish four different types of unresolved configurations:
\begin{itemize}
\item[(a)] One unresolved parton but the experimental observable selects only
$m$ jets;
\item[(b)] Two colour-connected unresolved partons (colour-connected);
\item[(c)] Two unresolved partons that are not colour connected but share a common
radiator (almost colour-unconnected);
\item[(d)] Two unresolved partons that are well separated from each other
in the colour chain (colour-unconnected).
\end{itemize}
Among those, configuration (a) is properly accounted for by a single tree-level three-parton antenna function like used already at NLO. Configuration (b) requires a tree-level four-parton antenna function (two unresolved partons emitted between a pair of hard partons), while (c) and (d) are accounted for by products of two tree-level three-parton antenna functions. On the other side, the one-loop single unresolved subtraction term $\d \sigma^{VS,1}_{NNLO}$ must account for three types of singular contributions:
\begin{itemize}
\item[(a)] Explicit infrared poles of the virtual one-loop
$(m+1)$ parton matrix element.
\item[(b)] Single unresolved limits of the  virtual one-loop
$(m+1)$ parton matrix element.
\item[(c)] Terms common to both above contributions,
which are oversubtracted.
\end{itemize}
For all these cases the detailed form of the subtraction term is given in~\cite{ourifant}. The only genuinely new ingredient appearing at NNLO are the four-parton initial-final antenna function $X^0_{i,jkl}$, which can be obtained by crossing the corresponding final-final antenna functions, and one-loop three-parton initial-final antenna function $X^1_{i,jk}$ which can be obtained by crossing from their final-final counterparts, listed in~\cite{ourant}. Both have to be integrated over the appropriate phase space.

\section{Integration of initial-final antenna functions at NNLO}
The initial-final antenna functions all have the scattering kinematics
$q + p_i \to p_1+p_2 (+p_3)$, where
\begin{displaymath}
\quad q^2 = -Q^2 <0\,, \quad p_i^2=0\,, \quad z=\frac{Q^2}{2\,q\cdot p_i} \,,\quad p_1^2=p_2^2=p_3^2 =0\,,
\end{displaymath}
and $p_3$ is present only for the NNLO real radiation antenna functions. Thus, integration over the final-state two-parton or three-parton phase space yields a result which depends only on $Q^2$ and $z$.

\begin{figure}[ht]
\begin{center}
\begin{minipage}[c]{.47\textwidth}
\centering
\includegraphics[width=1.0\textwidth]{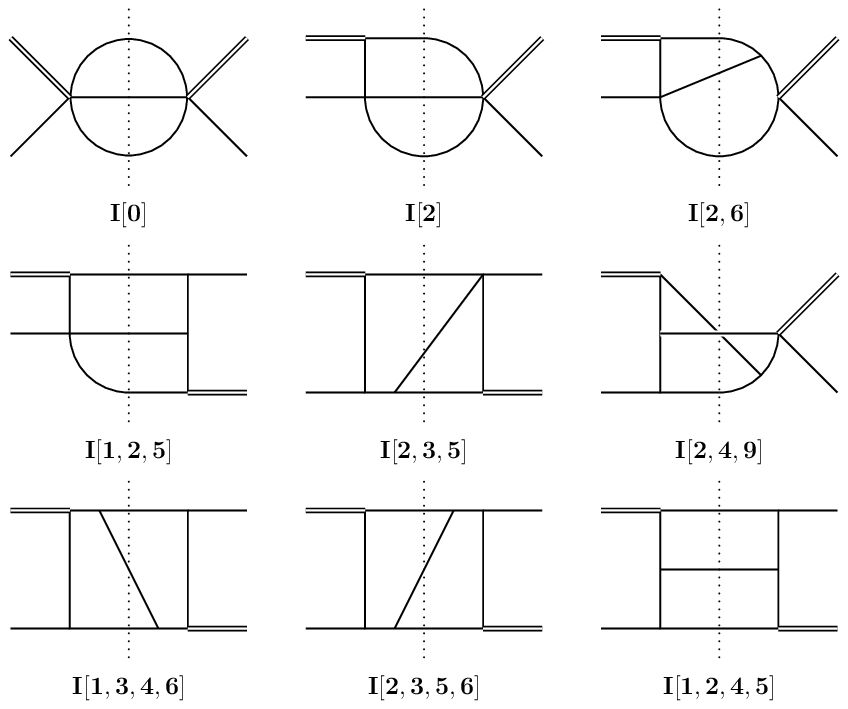}
\end{minipage}%
\hspace{5mm}%
\begin{minipage}[c]{.47\textwidth}
\centering
\includegraphics[width=1.0\textwidth]{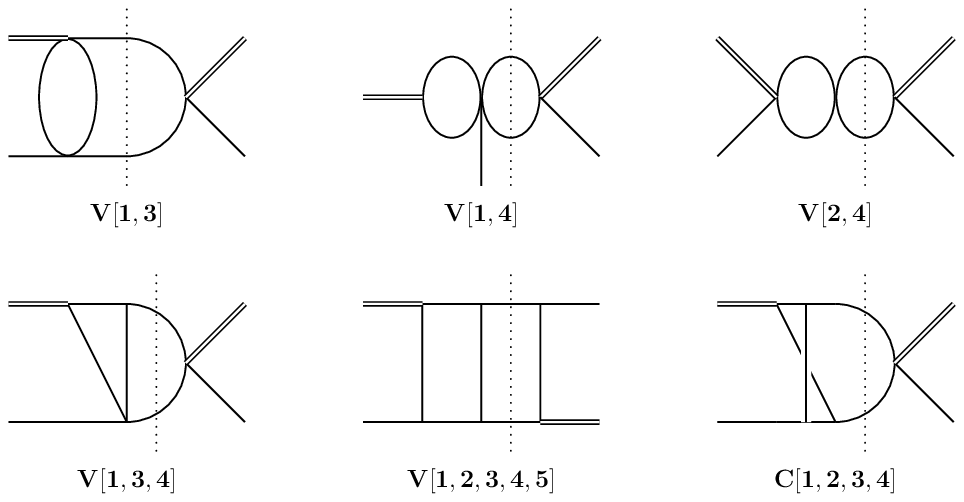}
\end{minipage}
\caption{Master integrals for the phase space integration of the double real tree level initial-final antennae at NNLO (left), and for the loop plus phase space integration of the one loop initial-final antennae at NNLO (right). The double line in the external states represents the off-shell momentum, $q$ with $q^2=-Q^2$, the single one is the incoming parton. All internal lines are massless. The cut propagators are the ones intersected by the dotted line.}\label{fig:masters}
\end{center}
\end{figure}

The NNLO double real radiation antenna functions $X^0_{i,jkl}$ have to be integrated over the inclusive three-parton final state phase space.  The NNLO one-loop single real radiation antenna functions $X^1_{i,jk}$ are integrated over the inclusive two-parton final state phase space, and over the loop momentum. For both types of integration, we first expressed all phase space integrals as loop integrals with cut propagators~\cite{babis}, and then employed the by-now standard technique of reduction to master integrals using integration-by-parts
(IBP,~\cite{chet}) and Lorentz invariance (LI,~\cite{gr}) identities among the integrals of any given topology. After carrying out the reduction, we found nine master integrals for the NNLO double real radiation antenna functions and six master integrals for the NNLO one-loop single real radiation antennae. All the masters integrals are shown in \figurename~\ref{fig:masters}. The convention for naming the master integrals follows the labelling of the numerators, i.e.
\begin{equation}
I[i,j,k] = \int\,
\frac{[\d p_1]\,[\d p_2]\,[\d p_3]}{D_i\, D_j\, D_k} \,
\delta^d(q+p_i - p_1-p_2 -p_3)\,,\quad\textrm{where}\quad [ \d p ]  =  \frac{\d^d p}{(2\pi)^d}  \delta^+ (p^2)\,.
\end{equation}
For the explicit definition of the propagators $D_{i}$ used in \figurename~\ref{fig:masters} we refer to~\cite{ourifant}.

All masters, except $I[1,2,4,5]$, have been computed by direct integration and by the differential equations method, supplemented, where necessary, by a direct calculation at $x=1$ after factorizing the leading singularity. Where appropriate, we compared our results to the expressions in the appendix of~\cite{zv}, finding full agreement. The explicit results of the master integrals up to the needed order in $\epsilon$ as well as the integrated initial-final antenna functions can be found in~\cite{ourifant}.

\section{Rederivation of NNLO coefficient functions}
Being derived from physical matrix elements, the integrated antenna functions can be compared to results from literature for inclusive cross
sections or coefficient functions, as was done previously for the final-final antennae in~\cite{our2j,ritzmann}.
In the case of the initial-final antennae, we can compare the integrated quark-antiquark antennae and gluon-gluon antennae against NNLO corrections to deep inelastic coefficient functions known in the literature. The former ones can be checked against DIS structure function
calculations~\cite{zv} whereas the latter can be compared to the $\phi$-DIS structure functions computed in~\cite{moch,moch2}. These structure functions are obtained in an effective theory with a scalar $\phi$ coupled to the square to the gluon field strength tensor. For example, the leading colour piece of the two-loop gluon initiated structure function can be written as the following linear combination of antennae:
\begin{equation}
\left.\mathcal{T}_{\phi,g}^{\l(2\r)}\right|_{N^{2}}=\,\mathcal{F}^{0}_{g,ggg} + 4\mathcal{F}^{1,R}_{g,gg} + 4\delta\l(1-z\r)\left.\l(2 F_{g}^{\l(2\r)} + F_{g}^{\l(1\r)\,2}\r)\right|_{N^{2}}\,,
\end{equation}
where $\mathcal{F}^{0}_{g,ggg}$ is the integrated tree level gluon-gluon double real radiation antenna, $\mathcal{F}^{1,R}_{g,gg}$ the integrated one-loop gluon-gluon antenna and $F_{g}^{\l(1\r)}$ and $F_{g}^{\l(2\r)}$ are respectively the one- and two-loop coefficients of the gluon form factor given in~\cite{formfactors}. An explicit expression for the two-loop quark- and gluon-initiated structure functions can be found in~\cite{ourifant,moch2}. The explicit linear combinations of antennae reproducing the different color contributions of the coefficient functions are given in~\cite{ourifant}.

The quark-gluon antennae, derived from neutralino decay, cannot be associated to any physical process and only the deepest pole structure could
be checked against a combination of Altarelli-Parisi splitting functions.

\section{Conclusions}
In this talk, we presented the extension of the NNLO antenna subtraction formalism~\cite{ourant} to include initial-final antenna configurations, where one of the hard radiator partons is in the initial state. Furthermore a highly non-trivial check of our results was performed by rederiving the two-loop coefficient functions for deep inelastic scattering. The subtraction terms presented here allow the construction of a parton-level event generator program for the calculation of NNLO corrections to jet production observables in deeply inelastic electron-proton scattering. Moreover, the initial-final antenna functions derived here are an important ingredient to the calculation of NNLO corrections to jet observables at hadron colliders, which will be possible once the computation of the initial-initial antenna configurations will be accomplished~\cite{radja}.

\section{Acknowledgments}
This research was supported in part by the Swiss National Science Foundation (SNF) under contracts PP0022-118864 and 200020-126691.

\end{document}